\documentclass{PoS}

\title{Trumpeting the Vuvuzela: \\ UltraDeep HI observations with MeerKAT}

\ShortTitle{Ultradeep HI observations with MeerKAT}

\author{Benne Willem Holwerda%
        \thanks{The authors are co-principal investigators on the MeerKAT UltraDeep HI Survey Key Program proposal.}\\
	Astrophysics, Cosmology and Gravity Centre (ACGC), \\
	Astronomy Department, University of Cape Town\\
       E-mail: \email{holwerda@ast.uct.ac.za}}
       \author{Sarah-Louise Blyth\\ 
	Astrophysics, Cosmology and Gravity Centre (ACGC), \\
	Astronomy Department, University of Cape Town\\
       E-mail: \email{sarblyth@ast.uct.ac.za}}

\abstract{The MeerKAT UltraDeep HI Survey aims to observe the 21 cm emission line of neutral hydrogen gas out to a redshift of z=1 and beyond. From both direct detections and stacked signal, we will address the \hi \ mass function, the cosmic neutral gas density of the Universe (\om) and their evolution over cosmic times, as well as galaxy evolution via e.g., the Tully-Fisher relation, the relation between \hi \ mass and Hubble Type or stellar mass, and the Schmidt-Kennicutt star-formation law. We propose to observe two fields, the COSMOS and Chandra Deep Field South (CDF-S) for 1000 hours each, adding an additional 4000 hours to one of these fields in 2015 when the full instantaneous bandwidth of MeerKAT (0.58-2.5 GHz) will be realised.}

\FullConference{ISKAF2010 Science Meeting \\
		 June 10 -14 2010\\
		 Assen, the Netherlands}
\def\hi{H\,{\sc i}}
\def\himf{H\,{\sc i}MF}
\def\om{$\Omega_{HI}$}

\usepackage{sidecap}
\usepackage{natbib}
\usepackage{multicol}
\usepackage{url}
 \usepackage{url}

\begin{document}

The MeerKAT radio telescope \citep[Karoo Array Telescope][]{meerkat1,meerkat2,MeerKAT}, a precursor instrument for the Square Kilometer Array \citep[SKA,][]{ska} is currently under construction in the Karoo, South Africa
The planned large bandwidth and high sensitivity will make MeerKAT the ideal instrument for high-redshift \hi \ observations until SKA is built. 
The proposed MeerKAT Ultradeep \hi \ Survey is designed to make optimum use of MeerKAT at each construction phase, in combination with existing surveys and a dedicated spectroscopic redshift survey with the Southern African Large Telescope.

\begin{figure}[htbp]
   \begin{minipage}{0.49\linewidth}
	\begin{center}
 	\includegraphics[width=\textwidth]{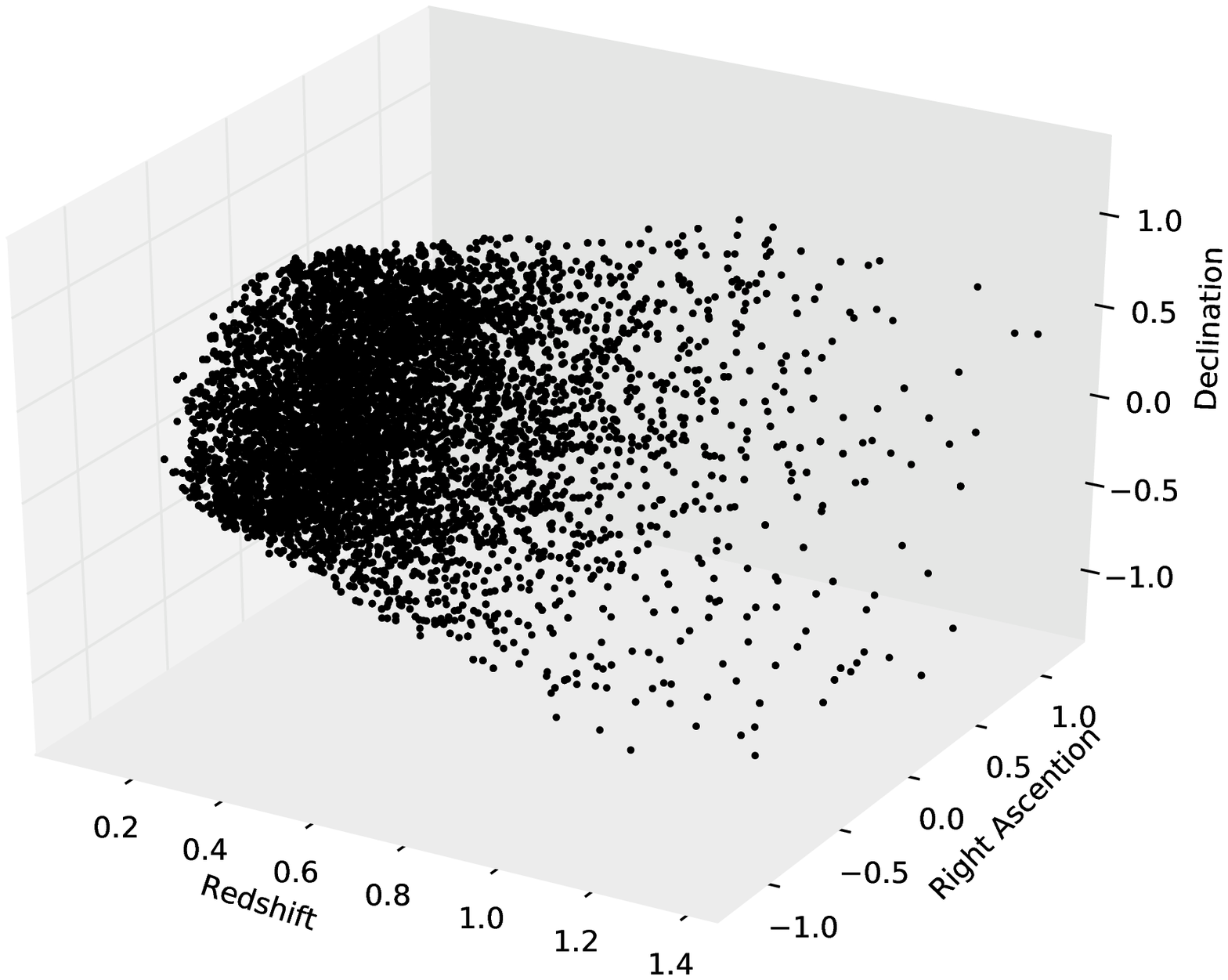}
	\caption{\label{f:vuv} A three dimensional representation of the Tier II datacube (5000 hours total). The field-of-view widens with redshift (z) resulting in the characteristic vuvuzela shape ($D=0.8^\circ$ at z=0). }
	\end{center}
   \end{minipage}\hfill
   \begin{minipage}{0.49\linewidth}
	\begin{center}
	\includegraphics[width=0.7\textwidth]{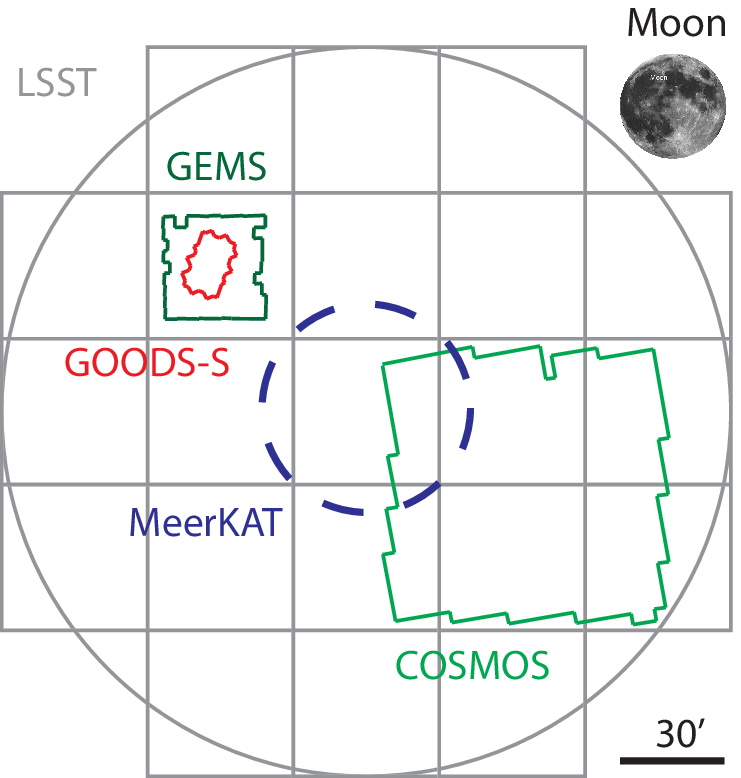}
	\caption{\label{f:fov}The field-of-fiew of the MeerKAT (at z=0) with those of the Large Synoptic Synthesis Telescope (LSST), approximately that of MeerKAT at z$\sim$1, and the Hubble Space Telescope deep fields of GEMS and GOODS on the CDF-S and COSMOS. }
	\end{center}
  \end{minipage}
\end{figure}

\section{Observations: the Vuvuzela}

The primary beam width of the 12m MeerKAT dishes increases with frequency and hence with redshift. As a result, a single pointing observation will cover a wider field-of-view at higher redshift, leading to a characteristic trumpet-like shape of the three-dimensional data volume, similar to a ``Vuvuzela''\footnote{The now well-known trumpet-like instrument used by soccer fans at the 2010 World Cup.}.
Figure \ref{f:vuv} gives an indication of the number and distribution of well-detected sources in the final 5000 hour combined observation.

\begin{figure}[htbp]
   \begin{minipage}{0.49\linewidth}
	\begin{center}
	\includegraphics[width=\textwidth]{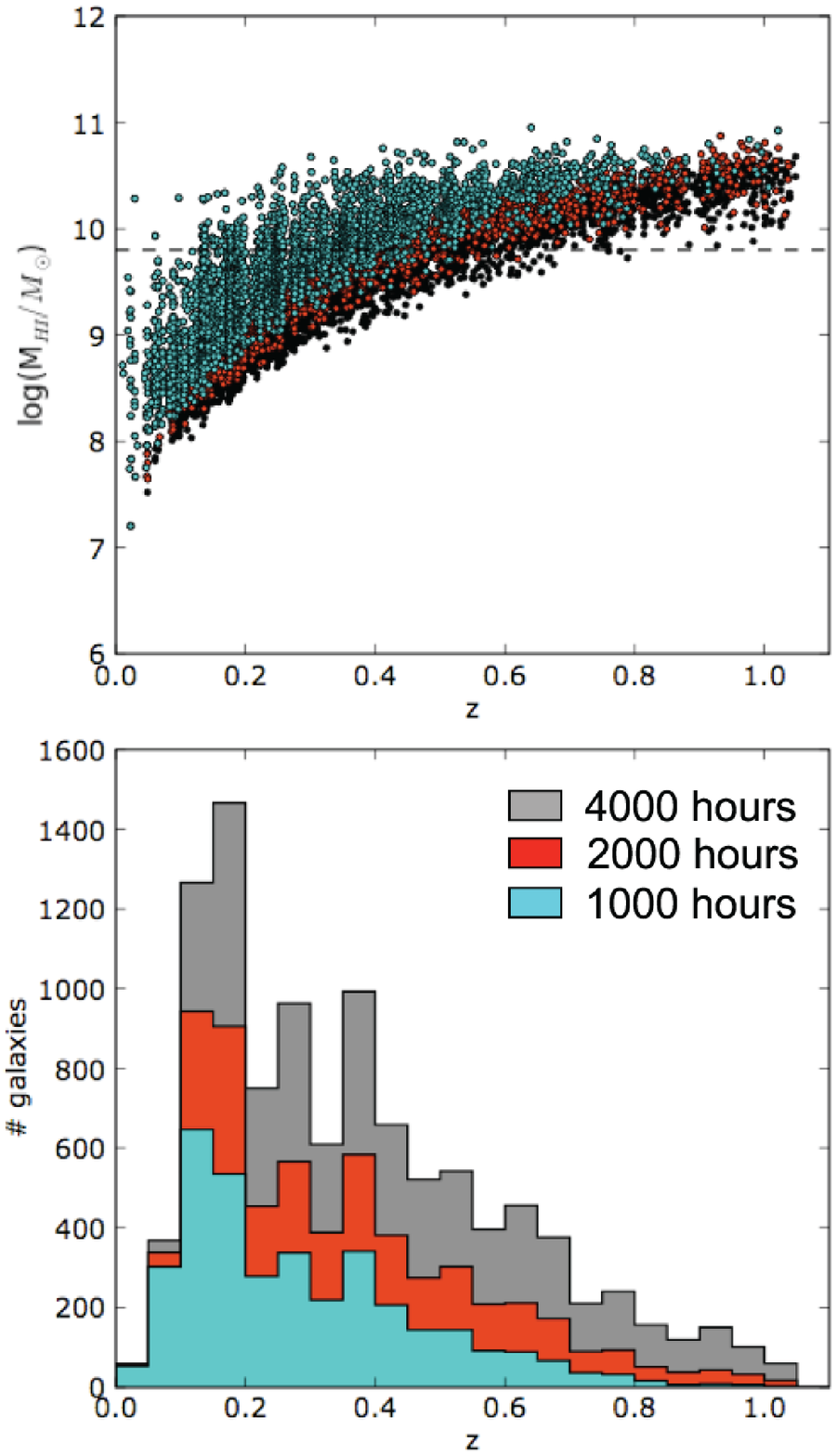}
	\caption{\label{f:counts} The 5$\sigma$ detections for three different integration times as a function of redshift. Top panel shows \hi mass versus redshift. The dashed line is $M_{HI}^*$, our target for direct detections with the ultradeep observations (Tier II), }
	\end{center}
   \end{minipage}\hfill
   \begin{minipage}{0.49\linewidth}
	\begin{center}
	\includegraphics[width=\textwidth]{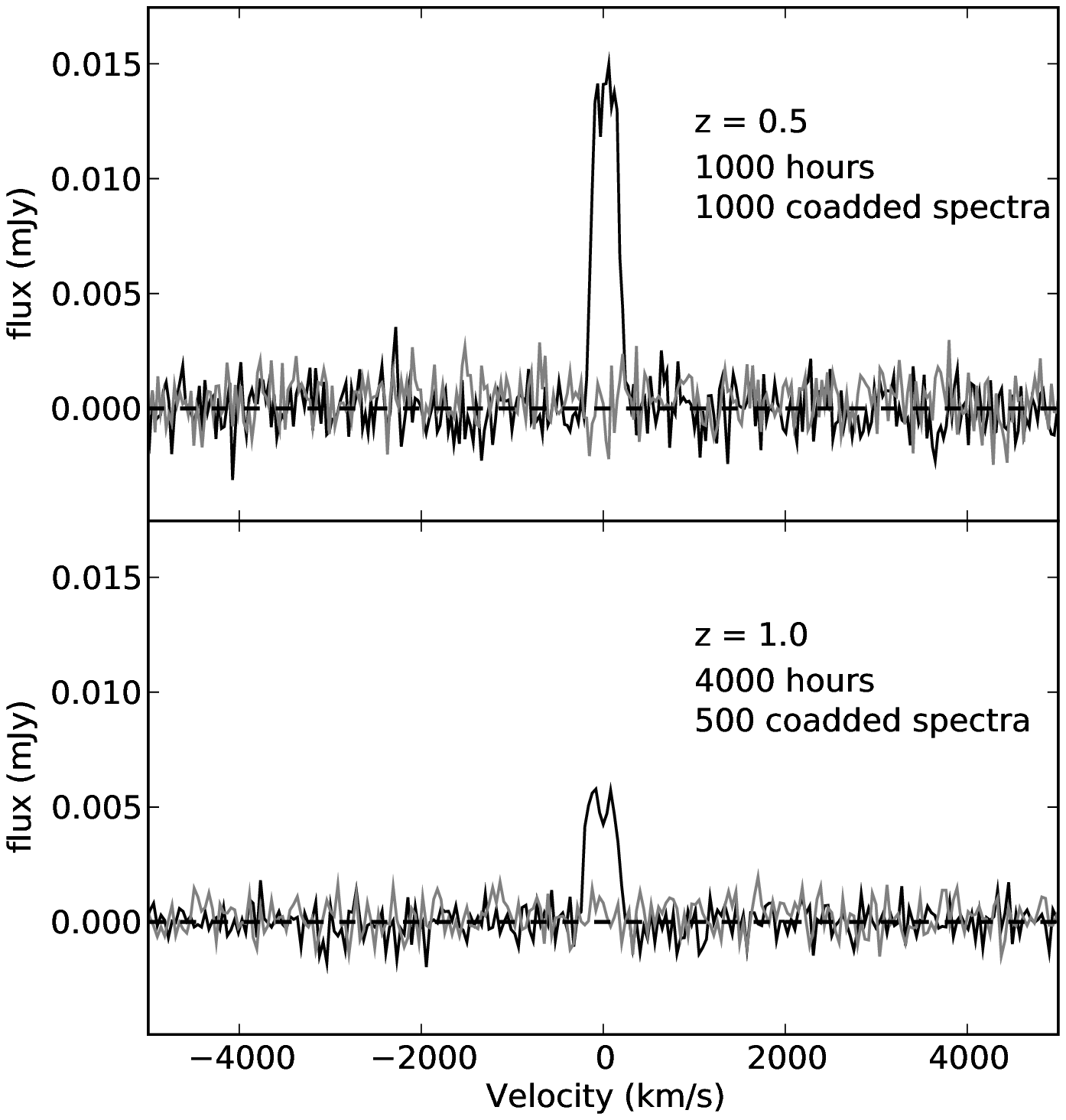}
	\caption{	\label{f:stack} Simulated stacking results (black lines) for z=0.5 (top panel) and z=1.0 (bottom panel), ($\Delta z$=0.1). The grey lines are reference spectra created by stacking the input spectra after shifting them by random redshifts. Galaxies were simulated according to the Oxford $S^{3}$ database \citep{Obreschkow09f}. For each redshift bin, the number of stacked spectra corresponds approximately to the currently available numbers of spectroscopic redshifts for the zCOSMOS survey.}
	\end{center}
   \end{minipage}
\end{figure}

We propose to do science with both direct detections and stacked spectra of objects. In the latter case, galaxy \hi \ spectra are shifted to the same reference frame using the known positions and redshifts of these objects and then co-added \citep[similar to][]{Verheijen07,Lah07,Lah09}. Combined, the signal-to-noise can be increased to yield an average line strength and width for these objects (Figure \ref{f:stack}). Different science questions can be explored this way but an existing spectroscopic redshift database is essential for successful stacking analysis. 

The observational strategy is dictated by the proposed science case and the roll-out of the MeerKAT construction (Table \ref{t:obs}), notably the expansion of available bandwidth after 2015. The choice of fields is guided  predominantly by the availability of spectroscopic redshifts as well as high-quality multi-wavelength data (Figure \ref{f:depth}). Because of the much larger field-of-view of MeerKAT compared to other wavelength deep surveys, these observations are effectively an \hi \ component to a wealth of multi-wavelength observations at the center, surrounded by a blind, ultradeep \hi \ survey (Figure \ref{f:fov}).

\begin{table}[h]
\caption{The goals of both observation phases of the MeerKAT Ultra-Deep \hi \ Field.}
\begin{center}
\begin{tabular}{l c c}
Survey Phases						& Tier I			& Tier II \\	
								& (2013-2015)		& (2016 - )\\	
\hline
\hline
\textbf{MeerKAT specs:}         			&                                  	&  \\
Bandwidth (GHz)      					&  0.9 - 1.75             	& 0.58 - 2.5 \\
Redshift range (z)						& 0.0 - 0.58		& 0.0 - 1.4  \\
\hline
\textbf{Survey Parameters} 		&                         		&         \\ 
Fields							& 2				& 1		\\
Observing time (hours)				& 2$\times$ 1000 h 	& +4000 h	\\
Spectroscopic redshifts 				&				&		\\
currently available:   					& 				& \\
\ \ full redshift range					& $\sim$10000 		& $\sim$1000  \\
\ \ highest redshift bin				& $\sim$1000 (at z=0.6)	& $\sim$ 500 (at z=1)        \\ 
redshift limits for:					&				& \\
Direct Detection of $M_{HI}^*$  		& z=0.4			& z=0.6 \\
\om \	using stacking					& z=0.6			& z=1.0\\
\hline
\end{tabular}
\end{center}
\label{t:obs}
\end{table}%

%

\section{Science}

Our Proposed key topic of investigation is galaxy evolution over cosmic time. OUr headline goals are therefore to measure the distriution of neutral hydrogen in galaxies, the \hi \ mass function (\himf) and the cosmic neutral gas density (\om) as a function of redshift. The \himf \ has to date only been determined for z=0 \citep{Zwaan05}, while the relation between \om \ and redshift is still ill constrained \citep{Lah07,Lah09,LahPhD}.

With the Tier I observations, we expect to observe galaxies with masses down to $M_{HI}^*$ out to redshift z=0.4 and to measure \om, using stacking, out to z=0.6, the limit of the initial bandwidth. In Tier II, we aim to observe $M_{HI}^*$ galaxies out to z=0.6 and anticipate that stacking will allow us to get an estimate of \om out to z$\sim$1, depending on size of the spectroscopic redshift catalog and noise characteristics of the MeerKAT.
%


Our secondary goal is to explore the evolution of galaxies though the \hi \ line, aside from the \himf. The relation between stellar mass, Hubble type or stellar bar, and the \hi \ content of galaxies as a function of look-back time can all be explored using both direct detections and stacked results. 

The wealth of multi-wavelength data, as well as the radio continuum, provide us with an estimate of the star-formation rate in these galaxies. The relation between gas-density and star-formation, the Schmidt-Kennicutt law, would need additional information on the molecular gas component in these galaxies. Atacama Large Millimetre Array (ALMA) observations would constitute an ideal complement for individual detections. Star-formation changes dramatically from z=1 to the present time \citep{Madau98,Hopkins06}, and the balance between and atomic and molecular hydrogen is the missing component to understanding the physics of the formation and evolution of disks over this time \citep{Obreschkow09a, Obreschkow09b, Obreschkow09c}.

\begin{SCfigure}
 	\includegraphics[width=0.5\textwidth]{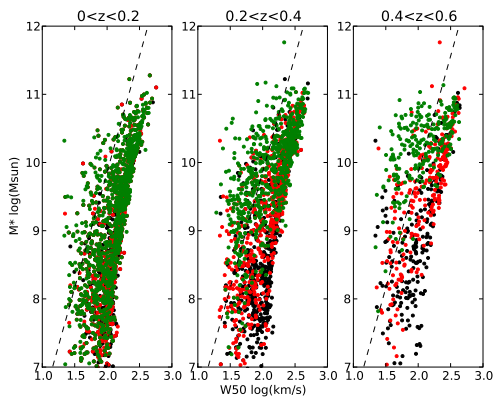}
	\caption{\label{f:tf} The Tully-Fisher relation: line width ($w50$) vs. stellar luminosity (mass), in three redshift intervals for solid detections (peak s/n $> 5\sigma$) for three integration times; 1000 (green), 2000 (red), and 5000 (black) hours, based on the $S^3$ SAX catalogue \citep{Obreschkow09f}. The MeerKAT UltraDeep  \hi \ observations will be able to study the slope, scatter, and normalization of the Tully-Fisher relation over a wide redshift range in great detail. }
\end{SCfigure}

The Tully-Fischer relation between line width and luminosity (or stellar mass), can be explored using directly detected galaxies (Figure \ref{f:tf}). The slope, scatter and normalization of this relation all depends on how rotationally supported disk galaxies assemble over the age of the Universe. We expect to dramatically increase the accuracy and the kind of T-F measurement at high redshift. The benefit of an \hi \ linewidth is that the atomic hydrogen disk probes the rotation curve well out to the point at which it flattens; the dynamics probe the whole halo mass for these galaxies. 
%


Additional science are the serendipitous detection of neutral gas in the cosmic web or ``dark", \hi-only galaxies, an accurate count of OH megamasers, and a comparison between the distribution of  \hi \ emission and absorption. 

%


\section{Complementary Data}

\begin{SCfigure}
\includegraphics[width=0.5\textwidth]{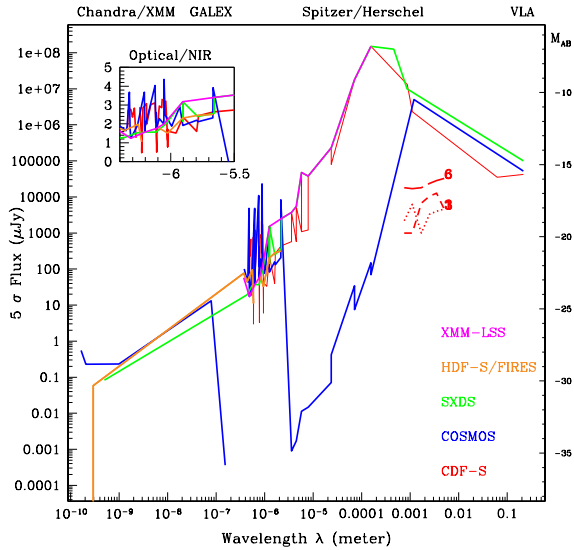}
\caption{\label{f:depth} The depth (5$\sigma$ flux level or limiting magnitude) as a function of wavelength for five deep fields accessible in the Southern Hemisphere. The COSMOS field stands out because of the low detection limits in the far-infrared but the Herschel Space Observatory is expected to obtain the deepest images on the CDF-S (dashed red lines, three different tiers of observations, 1, 3 and 6). These new observations will have better resolution, bringing them in line with the MeerKAT resolution, as well as reach several magnitudes deeper than the COSMOS FIR observations. The Large Synoptic Survey Telescope will improve limits in the optical SDSS filters by several magnitudes in its deep fields.}
\end{SCfigure}

The COSMOS and CDF-S fields have been observed across a broad range of wavelengths in great detail\footnote{see e.g., \url{http://www.strw.leidenuniv.nl/~jarle/Surveys/DeepFields/index.html}}. Figure \ref{f:depth} shows the limiting depth of observations as a function of wavelength. 
Both fields have ongoing spectroscopic redshift campaigns \citep{zcosmos,Balestra10}. Yet, because the multi-wavelength data does not cover the entire MeerKAT field-of-view, additional preparatory and follow-up observations will be necessary. Spectroscopic confirmation of the most distant and massive \hi \ lines will be paramount to removing contamination from OH megamasers. The stacking results from the Tier II field will also improve with a larger accurate redshift catalogue.
CO observations by ALMA will be needed to provide the molecular component of these distant galaxies. 

Therefore, we anticipate a substantial observational effort ($\sim$300 hours), with the Southern African Large Telescope (SALT) to generate a redshift catalogue in advance of Tier II observations, as well as follow-up observations with ALMA, and a deep optical field with SkyMapper and subsequently LSST. 

\section{Concluding remarks}

The MeerKAT UltraDeep \hi \ survey (MUDHI\footnote{We are open to suggestions for a better acronym.}), will revolutionalise \hi \ astronomy. For the very first time, the atomic gas component of distant galaxies, individual and per population, will be accurately known since z$\sim$1. The balance of MeerKAT capabilities, and the significant investment in observing time, will make this deepest \hi \ field a first instance of real SKA-type science before the SKA is constructed. 

\begin{multicols}{2}
\bibliographystyle{apj} 
\bibliography{/Users/holwerda/Desktop/Bib/Bibliography}
\end{multicols}

\end{document}